\documentclass[aps,prb,preprint,showpacs]{revtex4}
\usepackage{graphicx}% Include figure files
\usepackage{dcolumn}
\usepackage{bm}% bold math
\usepackage{color}

\begin{document}

\title{An approximation to density functional theory for
an accurate calculation of band-gaps of semiconductors}

\author{L. G. Ferreira}
\email{guima00@gmail.com}
\affiliation{Instituto de
F$\acute{\imath}$sica, Universidade de S\~{a}o Paulo, CP 66318,
05315-970 S\~{a}o Paulo, SP, Brazil}
\author{M. Marques}
\email{mmarques@ita.br}
\author{L. K. Teles}
\email{lkteles@ita.br}
\affiliation{Instituto Tecnol\'{o}gico
de Aeron\'{a}utica, 12228-900 S\~{a}o Jos\'{e} dos Campos, SP, Brazil}

\date{\today}

\begin{abstract}
The local-density approximation (LDA), together with the half-occupation (transition state) is
notoriously successful in the calculation of atomic ionization potentials. When it comes to
extended systems, such as a semiconductor infinite system, it has been very difficult to find a
way to half-ionize because the hole tends to be infinitely extended (a Bloch wave).
The answer to this problem lies
in the LDA formalism itself. One proves that the half-occupation is equivalent to introducing
the hole self-energy (electrostatic and exchange-correlation) into the Schroedinger equation.
The argument then becomes simple: the eigenvalue minus the self-energy has to
be minimized because the atom
has a minimal energy. Then one simply proves that the hole is localized, not infinitely
extended, because it must have maximal self-energy. Then one also arrives at an equation
similar to the SIC equation, but corrected for the removal of just 1/2 electron.
Applied to the calculation of band gaps and effective masses, we
use the self-energy calculated in atoms
and attain a precision similar to that of GW, but with the great advantage that
it requires no more computational
effort than standard LDA.
\end{abstract}

\maketitle
\section{Introduction}

The well known {\em Density-Functional Theory}
(DFT) \cite{KS} is an approach to the theory of electronic structure
 in which the electron density distribution, rather than the many electron
wavefunction, plays a central role. The practical applications of
DFT are based on approximations for the so-called
exchange-correlation potential which describes the effects of the Pauli
principle on the many-electron system.
If we had the exact exchange-correlation potential we could solve
the many-body problem exactly for the ground state.
Although the potential is unknown, approximations are made. The most
common is the so-called {\em Local-Density Approximation} (LDA) which
locally uses the exchange-correlation energy density of a
homogeneous system.
The LDA to the Kohn and Sham
DFT \cite{KS}
is still one of the most reliable methods for condensed matter calculations,
 having successfully predicted and explained a wide range of
ground state properties in solid state physics and chemistry \cite{review-Jones}.
Lately, but very slowly, it
is being progressively abandoned in favor of the many GGA (Generalized
Gradient Approximations) \cite{PBE}.
However, while LDA and GGA have predicted many ground state properties with good
accuracy, the electronic properties such as band gaps are significantly smaller
than experiment. These discrepancies are caused by  the lack of the
discontinuity of the exchange-correlation potential \cite{review-Jones} in
going from the valence to the conduction band.
Several methods for overcoming these limitations have been proposed. One of them
is the GW approximation, in which one considers the energies of quasiparticles
and calculate the electron self-energy in terms of perturbation theory.\cite{GW,Schilfgaarde}
This procedure has been quite successful, achieving good accuracy, but it goes
beyond the DFT. Other procedures were also proposed, among them we can mainly
cite  the Self Interaction Correction
(SIC) \cite{PZ}, { the atomic SIC applied to solids (ASIC) \cite{Vogel,Filippetti,Pemmaraju},
 which is
perhaps the procedure closest to ours,} hybrid functionals \cite{Becke},
screened exchange (SX-LDA) \cite{Asahi}, the so-called exact-exchange approach
\cite{Stadele}, the well
known LDA+U \cite{LDA+U}, the work of Liberman \cite{Liberman}, and others.
Most of these approaches are computationally very demanding,
which prohibits their application to large systems of atoms.
\par { Slater half-occupation scheme \cite{Slater,Leite,Johnson} was very successful
for valence states}.
 One example is that one could obtain energies which were
comparable to the experimental ionization energies \cite{Leite}, though,
at that time,
good spin-polarized exchange-correlation approximations,
as those based on Ceperley and Alder, did not exist.
\cite{PZ} In order to illustrate to the reader  the quality
of the results that can be obtained, we present in Table \ref{IP} the first and
second ionization
potentials of 12 atoms, measured and calculated with LDA with 1/2 occupation.

\begin{table}[htbp]
\begin{small}
    \centering
    \caption {First and second Ionization Potentials (IP) for some atoms (eV).
These results were obtained with spin-polarization but assuming spherical
charge densities for ions and atoms. We used a code originally written
by S. Froyen, modified by N. Troullier and J. L. Martins, and modified
and maintained by A. Garcia.}
    \label{IP}

        \begin{tabular}{l|cc|cr}
            \hline
  \hline
   & \multicolumn{2}{c|}{First IP}&\multicolumn{2}{c}{Second IP}\\

Atom & Calculation & Experiment & Calculation & Experiment \\
  \hline \hline
%B&8.46&8.30&24.70&25.15\\
C&11.60&11.26&24.58&24.38\\
N&14.81&14.53&30.01&29.60\\
O&13.89&13.62&35.38&35.12\\
Al&5.94&5.99&18.97&18.83\\\hline
Si&8.19&8.15&16.30&16.35\\
P&10.44&10.49&19.80&19.73\\
S&10.57&10.36&23.25&23.33\\
Zn&9.70&9.39&18.65&17.96\\ \hline
Ga&6.00&6.00&20.83&20.51\\
Ge&7.99&7.90&15.88&15.93\\
As&9.90&9.81&18.63&18.63\\
%Se&9.96&9.75&21.29&21.19\\ \hline
%Cd&8.86&8.99&16.82&16.91\\
In&5.73&5.78&18.56&18.97\\ \hline\hline
    \end{tabular}
    \end{small}
\end{table}
\par Though the precision of the calculated results shown in the Table
is much better than the precision one reaches in the calculation of band gaps, either
by LDA or GGA, it has been difficult to find a way to make the ionization of 1/2 electron in
extended systems as crystals. Of course the problem is that a crystal
is described by means of Bloch waves and removing the population of just one Bloch
state is of no consequence. In this paper we
present a solution to this problem. We are specially concerned with the
calculation of band gaps
in semiconductors, for which we obtain calculated results that
 compare very favorably with experiment, and are not
computationally demanding. We report the results for fourteen semiconductors,
including the group II-VI, III-V and IV.
Our method is inspired in the LDA and in the half ionization but, at some point,
it has to be postulated.
The quality of the results and the ease with which they are obtained show
that our assumptions are very good. Now we develop our method which could be
properly named as LDA-1/2.

\section{LDA and half-ionization in solids}
Accepting the LDA as a valid approximation to the density functional (DF), the
 Janak theorem\cite{Janak} follows
\begin{equation}\frac{\partial E}{\partial f_{\alpha}}=e_{\alpha}(f_{\alpha})
\end{equation}
where $E$ is the total energy of the system, function of the occupation $f_{\alpha}$
of
the one-particle Kohn and Sham state $\alpha$. It is a well known fact that
the eigenvalue $e_{\alpha}(f_{\alpha})$ is almost precisely linear with the
occupation $f_{\alpha}$.\cite{Leite} Then integrating
\[\int_{-1}^0df_{\alpha}\]
between the ground state $(f_{\alpha}=0)$ and the ion $(f_{\alpha}=-1)$
obtain
 \begin{equation}E(0)-E(-1)=e_{\alpha}(-1/2)=
-\ Ionization\ Potential.\label{ioniza}\end{equation}
Taking another derivative
\begin{equation}
\frac{\partial e_{\alpha}}{\partial f_{\alpha}}=2S_{\alpha},
\end{equation}
where
\begin{eqnarray}
S_{\alpha}&=&\int\int d^3rd^3r^{\prime}\frac{n_{\alpha}(\vec{r})n_{\alpha}
(\vec{r}^{\ \prime})}
{\big|\vec{r}-\vec{r}^{\ \prime}\big|}+
\frac{1}{2}\int\int d^3rd^3r^{\prime}n_{\alpha}(\vec{r})
\frac{\delta^2E_{xc}}{\delta n(\vec{r})\delta n(\vec{r}^{\ \prime})}
n_{\alpha}(\vec{r}^{\ \prime})\\
&+&\int\int d^3rd^3r^{\prime}\frac{n_{\alpha}(\vec{r})}
{\big|\vec{r}-\vec{r}^{\ \prime}\big|}\sum_{\beta}f_{\beta}\frac{\partial n_{\beta}
(\vec{r}^{\ \prime})}{\partial f_{\alpha}}
+\frac{1}{2}\int\int d^3rd^3r^{\prime}n_{\alpha}(\vec{r})
\frac{\delta^2E_{xc}}{\delta n(\vec{r})\delta n(\vec{r}^{\ \prime})}
\sum_{\beta}f_{\beta}\frac{\partial n_{\beta}(\vec{r}^{\ \prime})}{\partial f_{\alpha}}
\nonumber\label{self}\end{eqnarray}
is named ``self-energy'', because of the first term in the right.
In LDA the functional derivatives become common derivatives times 
delta-functions. We maintain the functional derivative notation because
the final formulae can have extended use.
Because of the linearity of $e_{\alpha}(f_{\alpha})$ { \cite{Leite}}we may write
\begin{equation}
e_{\alpha}(-1/2)=e_{\alpha}(0)-S_{\alpha}\label{DE}
\end{equation}
and
\begin{equation} E(0)=E(-1)+e_{\alpha}(0)-S_{\alpha}\label{six}\end{equation}
Eq.~\ref{six} is quite surprising in its simplicity. The Eq. is telling us that,
to restore the ground state, with total energy $E(0)$, from an ion
with a hole at state $\alpha$ we add an electron whose energy is the eigenvalue
$e_{\alpha}(0)$ minus the hole self-energy. The self-energy is large when
the function is much localized as an atomic wavefunction, and is small and zero
when it is much spread as a Bloch function. Since the energy of the restored
ground state must be a
minimum the hole self-energy must be a maximum. Thus the
hole should be representable by a very localized wavefunction. {\em So far, this is the
first time we see a demonstration of the hole localization}, though this proof is
based on an approximation (LDA) to the DF theory { and on the linearity assumption}. 
{ Of course we cannot say that the localized
hole state is truly stationary, specially if its energy is inside the band continuum
of the Bloch states, into which the localized hole would be scattered.}
\par The self-energy may be thought as the Quantum Mechanical average of a
``self-energy potential''
$V_S(\vec{r})$ such that
\begin{equation}S_{\alpha}=\int d^3r n_{\alpha}(\vec{r})V_S(\vec{r})\label{nVS}\end{equation}
where $n_{\alpha}=\psi_{\alpha}^*\psi_{\alpha}$,
\begin{eqnarray}
V_S(\vec{r})&=&\int d^3r^{\ \prime}\frac{n_{\alpha}(\vec{r}^{\ \prime})}{|\vec{r}-
\vec{r}^{\ \prime}|}+\frac{1}{2}\int d^3r^{\ \prime}\frac{\delta^2 E_{xc}}
{\delta n(\vec{r})\delta n(\vec{r}^{\ \prime})}n_{\alpha}(\vec{r}^{\ \prime})\label{VS}\\
&+&\int d^3r^{\ \prime}\frac{\sum_{\beta}f_{\beta}\frac{\partial n_{\beta}
(\vec{r}^{\ \prime})}{\partial f_{\alpha}}}{|\vec{r}-
\vec{r}^{\ \prime}|}+\frac{1}{2}\int d^3r^{\ \prime}\frac{\delta^2 E_{xc}}
{\delta n(\vec{r})\delta n(\vec{r}^{\ \prime})}\sum_{\beta}f_{\beta}
\frac{\partial n_{\beta}
(\vec{r}^{\ \prime})}{\partial f_{\alpha}}\nonumber
\end{eqnarray}
that depends on the state $\alpha$.
From now on, in Eqs. such as \ref{VS}, we will not write the last two terms,
those depending
on the derivative of the wavefunctions with respect to the occupation $f_{\alpha}$.

\par { To derive Eq.~\ref{six} we assumed linearity, aside from the Janak theorem.
The linearity results when the Kohn and Sham eigenfunctions of the ground state are equal to those
of the ion, which is correct to a large extent.\cite{Leite} Coherently we may neglect the 
last two terms in Eqs. such as \ref{VS}, and make the difference $E(0)-E(-1)$, of two
minima, an extremum. Then we minimize (extremize) $e_{\alpha}(0)-S_{\alpha}$ as suggested by Eq.\ref{six}.
To do so we must write a variational expression that, upon minimization (extremization),
leads to a differential Eq. for the hole wavefunction $\psi_{\alpha}(\vec{r})$. {\em At this
point we must set clearly what we imply by the term ``hole''. From Eq.~\ref{six} we
see that we are adding an electron to a hole state of the ion. The hole state
might be in a valence or conduction band, the only requirement being that the state
is empty in the ion. Thus by a ``hole'' we mean ``an electron filling an empty state'' or 
``a particle excitation''. This particle excitation may be in the valence or in the conduction
band.} }
\par { The variational expression must be such that upon calculating it
with the solution $\psi_{\alpha}(\vec{r})$ of the differential Eq. it returns back
the average $e_{\alpha}(0)-S_{\alpha}$. Then we write it as
\begin{equation}
e_{\alpha}-S_{\alpha}=\left<\psi_{\alpha}\left|-\nabla^2-2\sum_I\frac{Z_I}{|\vec{r}-\vec{r}_I|}
+2\int\frac{n_0(\vec{r}^{\ \prime})}{|\vec{r}-\vec{r}^{\ \prime}|}d^3r^{\ \prime}
+\frac{\delta E_{xc}}{\delta n_0(\vec{r})}-V_S\right|
\psi_{\alpha}\right>\label{ealfa}\end{equation}
where $n_0(\vec{r})$ {\em does not} include the hole wavefunction $\psi_{\alpha}$ and it is
the number-density of the ground state.
In this case, the wavefunction $\psi_{\alpha}$ is to be interpreted as the hole state
to be filled with an electron to restore the ground state.
\par Performing
the extremization} and then inserting Eq.\ref{VS} we obtain the equation below with
the top entries of the brackets $\big\{\big\}$
\begin{eqnarray}\label{hole}&&\left[
-\nabla^2-2\sum_I\frac{Z_I}{|\vec{r}-\vec{r}_I|}
+2\int\frac{n_0(\vec{r}^{\ \prime})}{|\vec{r}-\vec{r}^{\ \prime}|}d^3r^{\ \prime}
+\frac{\delta E_{xc}}{\delta n_0(\vec{r})}\right.\\
&&\left.-\big\{\begin{array}{c}1\\2\end{array}\big\}\int\frac{
\psi_{\alpha}(\vec{r}^{\ \prime})^*
\psi_{\alpha}(\vec{r}^{\ \prime})}
{|\vec{r}-\vec{r}^{\ \prime}|}d^3r^{\ \prime}
-\big\{\begin{array}{c}1/2\\1\end{array}\big\}\int
\frac{\delta^2E_{xc}}{\delta n_0(\vec{r})
\delta n_0(\vec{r}^{\ \prime})}
\psi_{\alpha}(\vec{r}^{\ \prime})^*\psi_{\alpha}(\vec{r}^{\ \prime})d^3r^{\ \prime}
\right]\psi_{\alpha}(\vec{r})=
\lambda_{\alpha}\psi_{\alpha}(\vec{r})\nonumber\end{eqnarray}
If we insert Eq.~\ref{VS} before extremization we obtain the Eq.~\ref{hole} with the
entries at the bottom of $\big\{\big\}$. In this latter case,
the next to last term of the operator is exactly the term in the
SIC equation\cite{PZ}. Its effect is to exclude the electron
being considered from the Hartree interaction. The last term in the operator,
the exchange-correlation term, is very different from the corresponding SIC term,
 since it depends on
the whole density of the system and not only on the density of the $\alpha$ state.
 The SIC equation, Eq.~\ref{hole} with the bottom entries in $\big\{\big\}$ is
not what
we want because the eigenvalue $\lambda_{\alpha}=e_{\alpha}-2S_{\alpha}$ and not
$\lambda_{\alpha}=e_{\alpha}-S_{\alpha}$ as the half-ionization requires.
{ It is worth
mentioning that, in the calculation of band gaps, SIC over corrects and halving it
seems to be a better procedure.\cite{Pemmaraju}}

\par { Except for atoms, solving Eq.~\ref{hole} is very difficult, either with the top or
bottom entries in $\big\{\big\}$.
 One important problem is that the solutions of Eq.~\ref{hole} are not orthogonal.
The SIC solution for atoms is used in the ASIC method, which is excellently reviewed in ref. \onlinecite{Pemmaraju}.
In our case we proceed differently. We introduce a parametrized self-energy
potential and use a variational expression that is an extremum for variations in the parameter(s).
The first question to be answered is whether it is possible to define a unique self-energy potential that is
state-independent. In Table~\ref{ase} we show a study of how the atomic self-energy of many states vary
with the assumed self-energy potential. One sees that for $s$ and $p$ orbitals, the self-energy does not
vary much whether it is calculated with $s$, $p$, and even $d$ self-energy potentials. Of course the self-energy
potential that we will use is the one corresponding to the atomic orbital dominating the crystal energy bands
around the gap. In extreme cases we can define a self-energy potential that is angular momentum dependent.
\begin{equation}V_S=\sum_lV_{S,l}(r)\sum_{m=-l}^l|l,m><l,m|\label{ioper}\end{equation}
This possibility was explored in the case of diamond, as shown in the discussion of our results.}
\begin{table}[htbp]
{\begin{small}
    \centering
    \caption {Atomic self-energies (eV), Eq.~\ref{nVS}, for some third-row atoms and different
valence electrons, calculated with self-energy potentials (Eq.~\ref{atomicVS}) derived
from half-ionization of different states. The fact that the self-energy of a given
state has no important dependence on the self-energy potential, that is the entries
at each column for a given atom do not differ much, shows that, for practical purposes,
one can neglect the state-dependence of the self-energy potential. In the
table below, only the self-energies of d-states seem to be much dependent
on the way the self-energy potential is derived.}
    \label{ase}
        \begin{tabular}{l|c|c|c|c}
            \hline
  \hline
    \multicolumn{2}{c|}{half-ionized }&\multicolumn{3}{c}{Self-energy (eV)}\\
\multicolumn{2}{c|}{valence state}&\ \ \ \ s\ \ \ \ &\ \ \ \ p\ \ \ \ &\ \ \ \ d\ \ \ \ \\
  \hline \hline
Zn&s&3.63&&4.62\\
&d&4.62&&7.43\\ \hline
Ga&s&3.15&3.67&4.67\\
&p&3.44&3.02&3.91\\
&d&4.75&4.02&8.12\\ \hline
Ge&s&4.13&3.71&5.03\\
&p&3.70&3.35&4.22\\
&d&5.11&4.30&9.00\\ \hline
As&s&4.11&3.73&5.26\\
&p&4.26&3.93&4.73\\
&d&5.33&4.57&9.79\\
\hline\hline
    \end{tabular}
    \end{small}}
\end{table}

\par {Thus, assuming a self-energy potential that is state-independent,
we use the following variational expression.\cite{celular,Harris}}
\begin{eqnarray}
E[n,v,\rho]&=&K[n]-\int V[p]\rho+\frac{1}{2}\int V[\rho]\rho-\int V_S\rho\nonumber\\
&+&\int v(n-\rho)+E_{xc}[\rho]
\label{extreme}\end{eqnarray}
where $p$ is the proton number-density, $n=\sum_{\beta}f_{\beta}\psi_{\beta}\psi_{\beta}^*$
is the electron number-density made out of the squares of the wavefunctions, $f_{\beta}$ being
the occupation numbers, $v$ is
the Kohn-Sham potential, $\rho$ is the model number-density, $V_S$ is the given
parametrized self-energy potential, $K$ and $E_{xc}$ have their usual meaning of kinetic and
exchange-correlation of the Kohn-Sham DFT. The functional $V[\rho]$ is defined as
\begin{equation}V[\rho(\vec{r})]=2\int\frac{\rho(\vec{r}^{\ \prime})}{|\vec{r}-
\vec{r}^{\ \prime}|}d^3r^{\prime}\end{equation}
\par The functional $E$ is {an extremum} for variations
in any of the three functions $n$, $v$, and $\rho$~: \begin{enumerate}
\item $\delta E/\delta v=0$ leads to $\rho=n$
\item $\delta E/\delta \rho=0$ leads to
\begin{equation}v=-V[p]+V[\rho]-V_S+\delta E_{xc}/\delta\rho\label{ksv}\end{equation}
\item $\delta E/\delta n=0$ leads to Schroedinger Equations with potential $v$ and
eigenvalues $e_{\alpha}$, rewritten as.
\end{enumerate}
\begin{equation}\left[
-\nabla^2-2\sum_I\frac{Z_I}{|\vec{r}-\vec{r}_I|}
+2\int\frac{\rho(\vec{r}^{\ \prime})}{|\vec{r}-\vec{r}^{\ \prime}|}d^3r^{\ \prime}
+\frac{\delta E_{xc}}{\delta \rho(\vec{r})}-V_S(\vec{r})\right]\psi_{\alpha}(\vec{r})=
e_{\alpha}\psi_{\alpha}(\vec{r})\label{rewritten}\end{equation}

\par Using Eq.~\ref{ksv} to determine $\rho$ for given $v$ and $V_S$
 and solving the Schroedinger equations, find
\begin{equation}
E=\sum_{\beta}f_{\beta}e_{\beta}-\frac{1}{2}\int V[\rho]\rho+E_{xc}[\rho]-
\int \rho\frac{\delta E_{xc}}{\delta\rho}\label{LDA-05}\end{equation}
It must be understood that both $n$ and $\rho$ are number densities of $N$ electrons, not
$N-1/2$, but the eigenvalues correspond to a situation where 1/2 electron
is removed, if $V_S$ is well chosen.
\par We want to find band gaps by taking the difference of total energies due to
different occupations $f_{\alpha}$. Maintaining the Kohn and Sham potential $v$ and
the model number-density $\rho$, solution of Eq.~\ref{ksv}, the band gap becomes
a difference between eigenvalues $e_{\alpha}$. Now, because the total energy
is a variational functional, that is an {an extremum} for variations in $v$, resulting from
variations of the self-energy potential $V_S$, one should look for extreme
eigenvalue differences.
\begin{equation}
\frac{\delta(e_{\alpha}-e_{\beta})}{\delta V_S}=0\label{diferenca}\end{equation}

\section{The LDA-1/2 method}
Consider the case of an atom. We first prove that the self-energy potential is
given by
\begin{equation}V_S\simeq-V(-1/2,r)+V(0,r)\label{atomicVS}\end{equation}
namely the difference between the all-electron potentials of the atom and
of the half-ion.
\par We begin by writing the potential difference as
\begin{eqnarray}
V(-1/2,r)&-&V(0,r)=\int_0^{-1/2}df_{\alpha}\frac{\partial}{\partial f_{\alpha}}
\left\{-2\frac{Z}{r}+2\int d^3r^{\ \prime}\frac{n(\vec{r}^{\ \prime})}{|\vec{r}-
\vec{r}^{\ \prime}|}+\frac{\delta E_{xc}}{\delta n(\vec{r})}\right\}\nonumber\\
&=&\int_0^{-1/2}df_{\alpha}\left\{2\int d^3r^{\ \prime}\frac{n_{\alpha}
(\vec{r}^{\ \prime})}{|\vec{r}-
\vec{r}^{\ \prime}|}+\int d^3r^{\ \prime}\frac{\delta^2 E_{xc}}
{\delta n(\vec{r})\delta n(\vec{r}^{\ \prime})}n_{\alpha}(\vec{r}^{\ \prime})
\right\}\\
&+&\int_0^{-1/2}df_{\alpha}\left\{2\int d^3r^{\ \prime}\frac{\sum_{\beta}
f_{\beta}\frac{\partial n_{\beta}
(\vec{r}^{\ \prime})}{\partial f_{\alpha}}}{|\vec{r}-
\vec{r}^{\ \prime}|}+\int d^3r^{\ \prime}\frac{\delta^2 E_{xc}}
{\delta n(\vec{r})\delta n(\vec{r}^{\ \prime})}\sum_{\beta}f_{\beta}
\frac{\partial n_{\beta}
(\vec{r}^{\ \prime})}{\partial f_{\alpha}}\right\}\nonumber\label{mean}
\end{eqnarray}
or, for a certain value of $f_{\alpha}$ in $[-1/2,0]$
\begin{eqnarray}
-V(-1/2,r)&+&V(0,r)=
\int d^3r^{\ \prime}\frac{n_{\alpha}(\vec{r}^{\ \prime})}{|\vec{r}-
\vec{r}^{\ \prime}|}+\frac{1}{2}\int d^3r^{\ \prime}\frac{\delta^2 E_{xc}}
{\delta n(\vec{r})\delta n(\vec{r}^{\ \prime})}n_{\alpha}(\vec{r}^{\ \prime})
\nonumber\\
&+&\int d^3r^{\ \prime}\frac{\sum_{\beta}f_{\beta}\frac{\partial n_{\beta}
(\vec{r}^{\ \prime})}{\partial f_{\alpha}}}{|\vec{r}-
\vec{r}^{\ \prime}|}+\frac{1}{2}\int d^3r^{\ \prime}\frac{\delta^2 E_{xc}}
{\delta n(\vec{r})\delta n(\vec{r}^{\ \prime})}\sum_{\beta}f_{\beta}
\frac{\partial n_{\beta}
(\vec{r}^{\ \prime})}{\partial f_{\alpha}}
\label{dif}\end{eqnarray}

\begin{figure}[htbp]
\begin{center}
\includegraphics[scale=0.6]{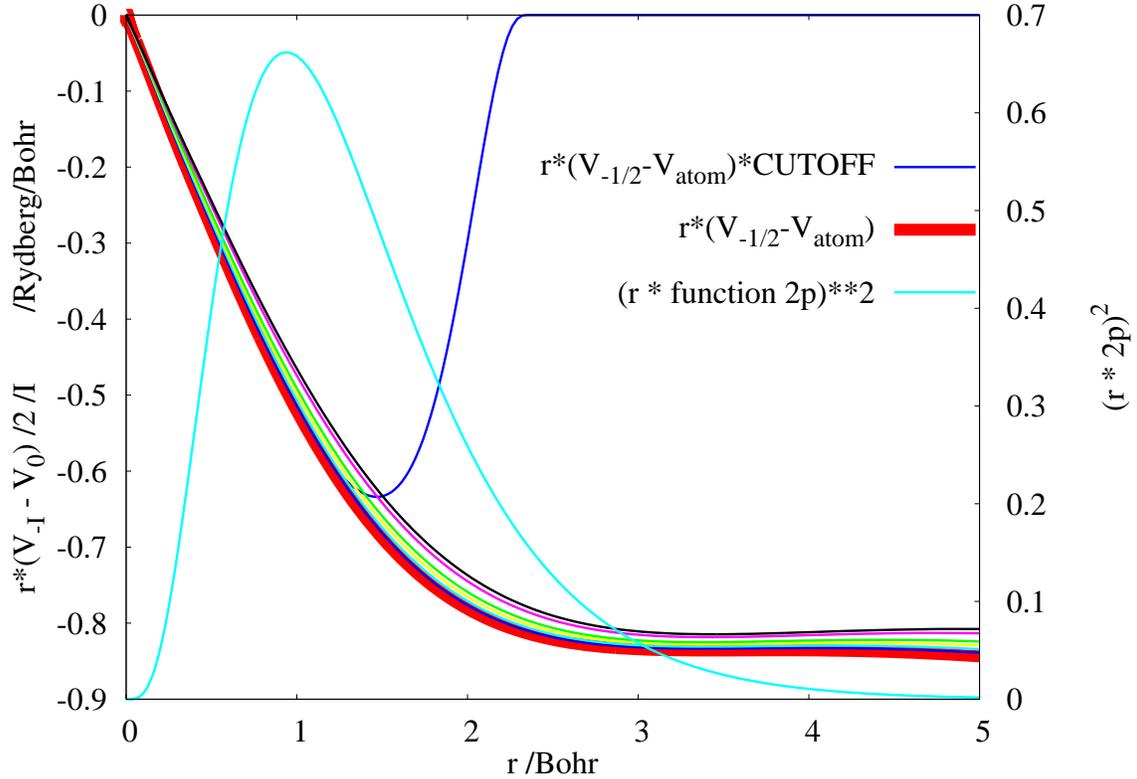}
\caption{\small{(Color online) Self-energy potential ($rV_S$) calculated for the N atom at different
ionizations $I$ ranging from $0.5$ to $-0.2$. The lines bunch around
that of $I=0.5$, which is made thicker. The wavefunction for the ionized state (2p) is
also shown.
Also shown is the potential after the cut-off by $\Theta(r)$.
}}
\label{05+02}
\end{center}
\end{figure}
\par  Fig.~\ref{05+02} depicts $r$-times the self-energy
potential for the nitrogen atom, a typical case,
for degrees of ionization $I$ ranging from $0.5$ to $-0.2$.
Observe that the ratio
\begin{equation}\frac{V(f_{\alpha},r)-V(0,r)}{f_{\alpha}}\end{equation}
has a very poor dependence on $f_{\alpha}$ meaning that in Eq.~\ref{dif} we can take
the $xc$ functionals at the full occupation $f_{\alpha}=0$.
Then, comparing Eqs.~\ref{VS} and \ref{dif} our proof is completed.

\par We will leave to another paper a discussion on the lone hole solution
we can get out of Eq.~\ref{hole}. Here we are interested in calculating band gaps
of semiconductors. For that purpose we will repeat the atomic self-energy
potential (Eq.~\ref{dif}) in the whole lattice and calculate eigenvalues for
``hole bands''. But observe that the first term on the right of Eq.~\ref{dif}, when
repeated in the whole crystal, diverges because it is Coulomb-like. On the
other hand, as Fig.~\ref{05+02} shows, the Coulomb tail of the atomic $V_S$
has no importance because the wave function never goes far.
Then, in using the self-energy potential $V_S$ defined in the atoms
 we first trim the potential with a function as
\begin{equation}\Theta(r)=\left\{\begin{array}{lr}
\left[1-\big(\frac{r}{CUT}\big)^n\right]
^3&r\le CUT\\
0&r>CUT\end{array}\right.\label{teta}\end{equation}
The idea is that the atomic self-energy potential is only meaningful where
the atomic wave function is not negligible. Of course the trimmed self-energy
potential is being repeated throughout the infinite crystal, so that we are
actually calculating ``filled hole bands''. Due to the
trimming, the Coulomb tail (of -1/2 electrons) of the atomic $V_S$ does not
penetrate into the neighboring atoms. With the trimming,  the
eigenvalues $e_{\alpha}$ in Eq.~\ref{rewritten} become dependent on the
trimming parameter $CUT$. However, Eq.~\ref{diferenca}  sets a recipe to
choose the value of $CUT$: one should make the energy gaps extreme.

\par
\begin{figure}[htbp]
\begin{center}
\includegraphics[scale=1.2]{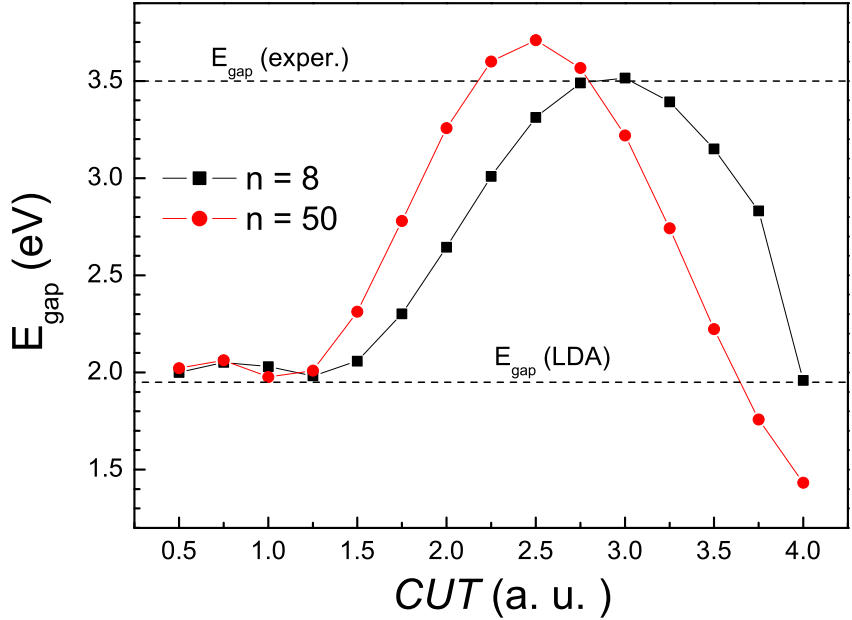}
\caption{\small{(Color online) Band gap of GaN as function of the parameter $CUT$
of the cut-off function $\Theta(r)$ applied to the self-energy
potential of {\em N 2p}.  The band gap extreme values depend
to some extent on the exponent $n$ being used (n=8 and 50 are shown).
}}
\label{f(cut)}
\end{center}
\end{figure}

 The function of Eq.~\ref{teta} has some important properties: 1 - its derivative is also zero at
$r=CUT$, so
that its electric field is zero at that point and the cut-off does not add to
the total
charge of the atom; 2 - the trimmed self-energy potential $\Theta(r)V_S(r)$
is wholly contained
inside a sphere of radius $CUT$, which facilitates its use in
band-calculation methods such
as SIESTA and APW -like. The power $n$ should be large so
that the cut-off is sharp. In actual
practice we tried $n=8$ and $n=50$ with similarly good results,
thus we adopted $n=8$, which is less abrupt, and does
not introduce numerical problems into the
programs. Fig.~\ref{f(cut)}
shows a typical behavior of a band gap as function of the parameters defining
the cut-off function. The first increase of the band gap with $CUT$ only means
that we are getting more of the valence band self-energy as the cut-off is
made at larger radii. In principle the larger the $CUT$ the more we get
of the valence self-energy. After reaching a
peak, the gap decreases because (i) the potential $V_S$ is penetrating into
neighboring atoms, tending to a uniform negative potential everywhere in space,
shifting downwards all bands, valence and conduction alike, and (ii) the self-energy
potential perturbation, being broad, diffuses the excitation wavefunction thus
making it to loose locality and self-energy. In other words, the cut-off function
should be broad enough so as to include most of the excitation wavefunction and
thin enough so as not spread it.
 Thus, the procedure to determine $CUT$ is based in Fig.~\ref{f(cut)}, namely we look
for the extreme band gap according to Eq.~\ref{diferenca}.

\par
\begin{figure}[htbp]
\begin{center}
\includegraphics[scale=1.5]{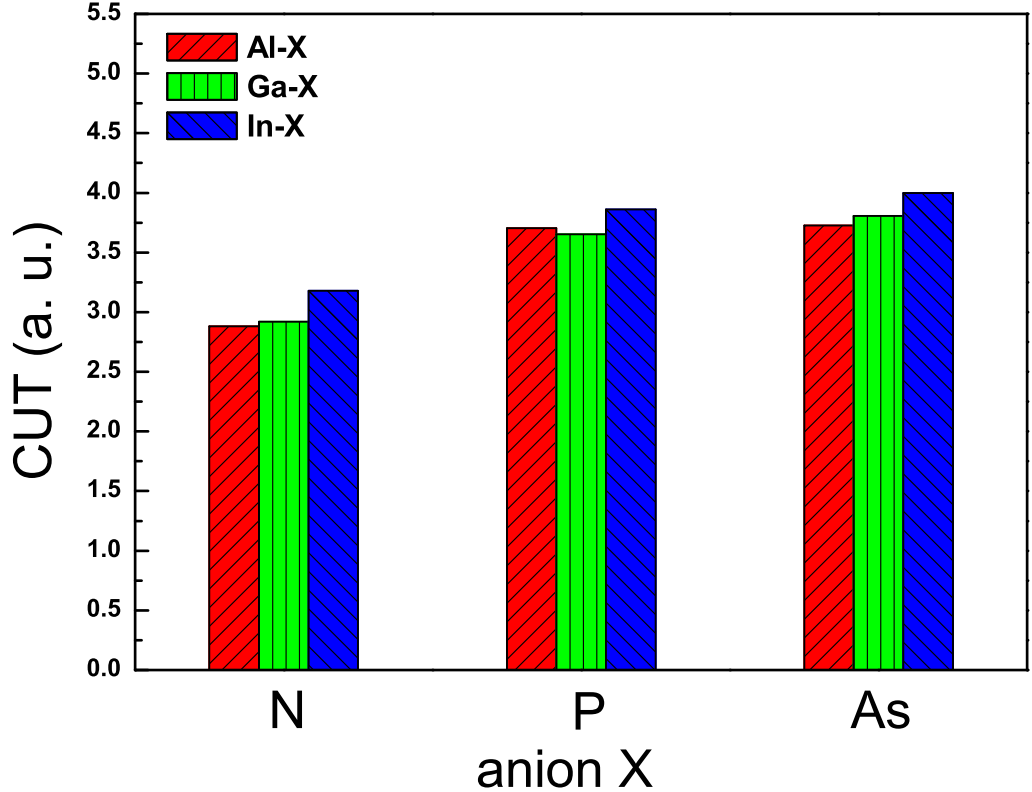}
\caption{(Color online) The optimized values of the parameter $CUT$, in which we correct
only the anion p-state for nitrides, phosphides and arsenides compounds.}
\label{cut}
\end{center}
\end{figure}

 {For a given atom and bonding type, the value of $CUT$ 
depends little on the chemical environment. Because we are using
$CUT$ values that make the gaps extreme, small deviations from
the optimal values produce only second-order deviations in the gaps.} 
In the Fig. \ref{cut} we present the values of anion p state CUT optimized for
arsenides, phosphides and nitrides. { The anion CUT value has a small dependence on the chemical environment, which is approximately linear with the compound bond length}. However, the relative variation produced in the energy gap is very small. This can be easily verified in the Fig. \ref{f(cut)}.  Around the energy gap maximum, the  range of CUT found in the optimization for all nitrides leads to a change of only 0.05 eV in the energy gap value. This behavior was also verified for all other calculated compounds. Therefore, we conclude that is very reasonable to consider the same CUT value for the anion potentials.  In the Table~\ref{raios} we show the optimal values of the parameter $CUT$
of the trimming function $\Theta$ of Eq.~\ref{teta}. The values for $CUT$ in the Table reminds a
table of ionic, covalent, or atomic radii, but are not equal to either.

\section{Results and Discussion}

We calculated, within the LDA-1/2 approach, the electronic structure for
several semiconductors. By comparing the LDA and LDA-1/2 calculation procedures,
the LDA-1/2 calculations lead to no more computational
effort than standard LDA, because the values of $CUT$ depend only on
the atoms, not on their environment,
 and are calculated just once. This is a great advantage of our method.
Most of the calculations were made with the code VASP ``Vienna {\it Ab-initio}
Simulation Package'' using the ultra-soft
pseudopotential\cite{Vanderbilt,Kresse} to which we added the trimmed self-energy
potential.
In some instances we
repeated the calculations
with the SIESTA code \cite{Soler} and the results differed by no
more than $0.1eV$. The
two codes are so different, for they use different basis
functions and pseudopotentials,
that the agreement of their results runs in favor of the reliability of our LDA-1/2
procedure. The ${\bf k}$-space integrals were approximated by
sums over a 9$\times$9$\times$9 special-point of the
Monkhorst-Pack type  within the irreducible part of the Brillouin
zone.\cite{DJ-73}  The number of plane waves for the expansion
of wave functions was optimized for each system and it is
basically the same value obtained for optimization of the equivalent
standard LDA calculation.  The lattice parameters used were the experimental ones.

\par  The present LDA-1/2 proposal is assuming that in
promoting an electron
from the valence band to the conduction band, the hole thus created is similar to
the hole created in the atomic photo-ionization. In other words, the hole has
the extent of an atomic hole.
If the hole in the extended system overlapped $N$ equal atoms its
self-energy would be $1/N^2$ that of the atomic hole, and the self-energy potential
$V_S$ would be $1/N$ that of the atom, or it would have to be calculated with
an ion with $1/2N$ electrons removed. The results for the semiconductors
III-V and II-VI, to be presented shortly, definitely point to $N=1$, meaning that
the hole in the solid resembles much the hole in the atom. In fact, the valence
band of these semiconductors is known to be made of the anion
wavefunctions. On the other hand, for
the IV elements Ge and Si (and also for diamond), the results point to $N=2$
meaning that the hole covers the two atoms with covalent bond.

\begin{table}[htbp]
\begin{small}
    \centering
    \caption {Values of $CUT$ that make the band gaps extreme, that is, when the
self-energy potential is defined by Eq.~\ref{atomicVS} and trimmed by Eq,~\ref{teta}.
The optimal value of $CUT$, as is the case of an ionic or covalent radius, is
typical of each atom,
 and the orbital that was half-ionized. In most cases only the anion matters.}
    \label{raios}
        \begin{tabular}{lcr||lcr}
            \hline
  \hline
&Half-ionized&&&Half-ionized&\\
Atom&orbital&$CUT$ (a.u.)&Atom&orbital&CUT (a.u.)\\\hline
Si&p&3.67 &Ge&p&3.46\\
N&p&2.90&P&p&3.86\\
As&p&3.81 &Zn&d&1.665 \\
O&p& 2.67&S&p&3.39 \\
Ga&d&1.23&In&d&2.126\\
\hline
  \hline
    \end{tabular}
    \end{small}
\end{table}

\par The band gaps calculated with LDA-1/2 are presented in Table~\ref{results}.
Here we must
remind that LDA-1/2
is still a scheme to calculate excitations, not the total energy and the equilibrium
 lattice parameter.
 Whereas the LDA
results exhibit the well known underestimation of the energy gap, LDA-1/2 results
present an excellent agreement with experiment. In general, by comparing the
theoretical LDA and LDA-1/2 band structures we observe that, as in LDA-1/2 the
self-energy is removed, the valence states are now more localized and are pulled
down in energy in comparison with the LDA, which results in a larger energy gap.
\par { The LDA-1/2 entries in Table~\ref{results} require a further explanation.
In the cases marked with two asterisks ** we are adding the trimmed self-energy potential
derived from the half-ionized anion p-state, and the trimmed self-energy potential
derived from the half-ionized cation d-state. The questions then are why adding the
p-correction to the anion and not the s-correction, and why d and not s to the cation.
The case of C, Si and Ge, when we used -1/4 and not -1/2 ionization, has been discussed above.
Thus there seems to be a certain degree of arbitrariness in a LDA-1/2 scheme. But that is not so
because, from what is known from the chemical bonding of these compounds, we could not
proceed differently. Further we are always keeping in mind the criterion of
an extreme band gap (Eq.~\ref{diferenca}). The case of diamond (C) is even more puzzling
because we are adding trimmed self-energy s and p potentials to a single atom. In this
case we are defining the self-energy potential as in Eq.~\ref{ioper} and approaching
 the method of
Filippetti and Spalding \cite{Filippetti}.
Again, Eq.~\ref{diferenca} is our guide.}

\begin{figure}
\includegraphics[scale=0.7]{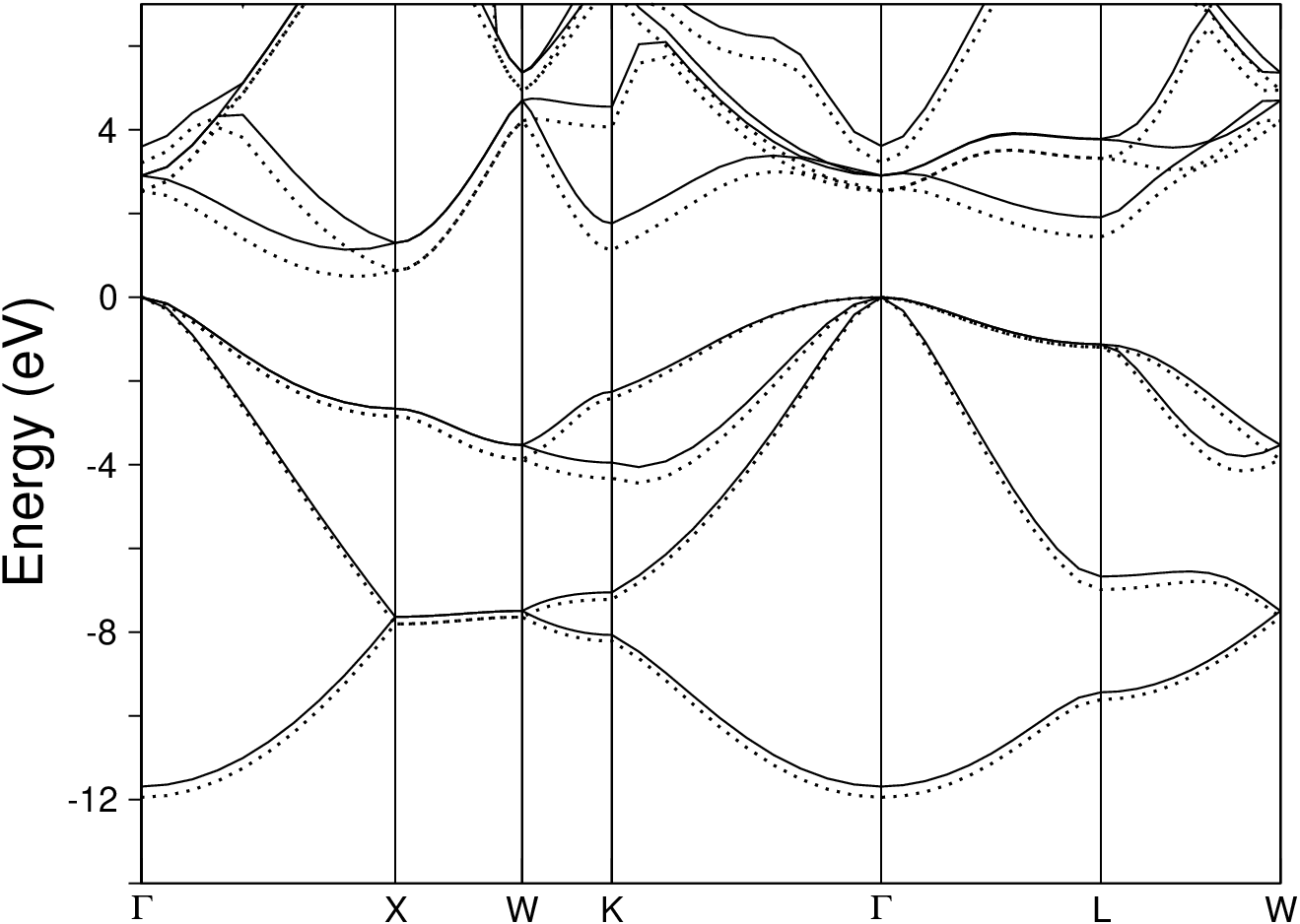}
\caption{Calculated band structures for Si (in eV). 
Dashed lines display LDA and solid lines represent LDA-1/2 results. 
The zero of energy was placed at the top of the valence band.}
\label{banda_Si}
\end{figure}

\begin{figure}
\includegraphics[scale=0.7]{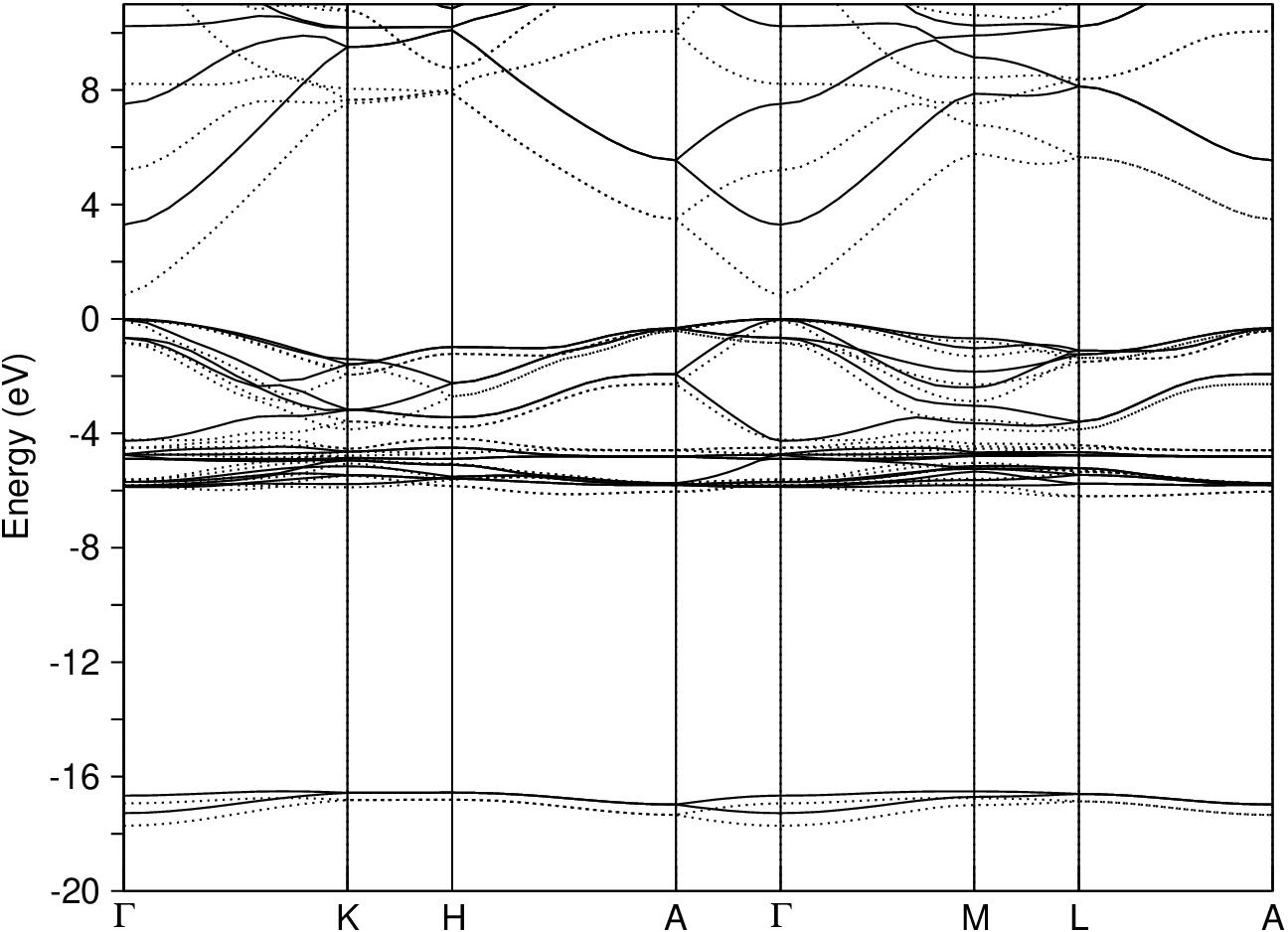}
\caption{Calculated band structures for ZnO (in eV). 
Dashed lines display LDA and solid lines represent LDA-1/2 results. 
The zero of energy was placed at the top of the valence band.}
\label{banda_ZnO}
\end{figure}

\begin{figure}
\includegraphics[scale=0.7]{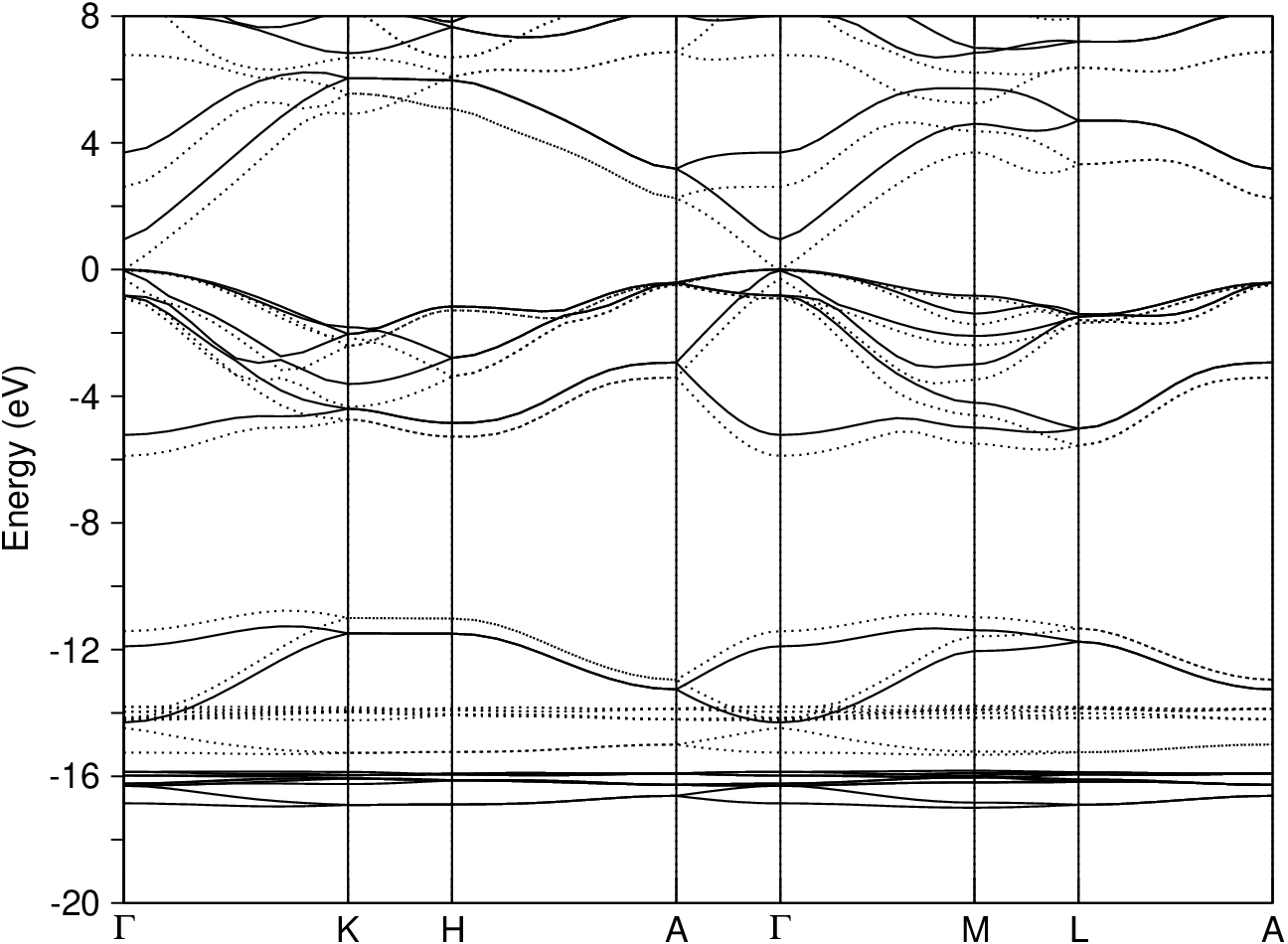}
\caption{Calculated band structures for InN (in eV). 
Dashed lines display LDA and solid lines represent LDA-1/2 results. 
The zero of energy was placed at the top of the valence band.}
\label{banda_InN}
\end{figure}

 Figs. \ref{banda_Si}, \ref{banda_ZnO} and \ref{banda_InN}  depict the
corresponding band structures (BS) along the main symmetry directions of
the Brillouin Zone (BZ) for Si, ZnO, and InN comparing the LDA-1/2 with LDA.
The zero of energy was placed at the top of the valence band. We chose to
show the BS for these semiconductors for two reasons: firstly, silicon is
one of the most important semiconductors, secondly, we would like to show the
results for cases where the LDA fails completely, as InN, for which LDA gives
a semi-metal instead of semiconductor, and ZnO, which became a
very interesting material with large band gap, and the LDA predicts an energy
gap much smaller than the experimental value. Moreover, for ZnO it
is  difficult to obtain the correct BS, even performing quasiparticle
calculations using GW if the starting point is the standard LDA
wavefunctions.\cite{MSchefler_2005}
From our results, we observe that for Si, the LDA-1/2 dispersion relations
are similar to the LDA but with the correct band gap energy. For the InN and
ZnO, the same behavior as Si occurs, but with some differences concerning
the cation d-states. In both InN and ZnO, the semicore cation d states
 play an important role. In the nitride, the states derived from
the atomic 4d~(In)-orbital lie close to the bottom of
the valence band 2s~(N)-like and hybridize with it \cite{Ramos_PRB}.
On the other hand, in ZnO, the cation d states lie approximately in the
middle of the valence band. Moreover, recently it was shown that in both
cases the d-states interact and hybridize with the top of valence band,
and in DFT-LDA there is an underestimation of the binding energies of these
semicore d states and consequently an overestimation of their hybridization
with the anion. The enhanced p-d coupling then pushes up the
valence-band maximum and the energy gap becomes 
smaller. \cite{SHWei_PRB_88,SHWei_PRB_2003,VandeWalle_APL_97,Janotti_PRB06}
By taking these facts into account, for InN and ZnO, LDA-1/2 corrects not only the
top of valence but also the cation d-states. Thus, by comparing the LDA with
LDA-1/2 band structures, we observe that the latter has the cation d-orbitals
deeper in energy. This effect is more pronounced in InN. In both cases,
ZnO and InN,  LDA-1/2 is
remarkable, leading to values very near the experimental ones.
In order to study deeply the influence of the cation d-state in both InN and
ZnO, in the Table \ref{VBW} we present the results for the valence band width (VBW)
and energy gap at different levels of the LDA-1/2 procedure. 
We observe that in the case
of ZnO, the O p-state correction increases the VBW and the Zn d-state correction
decreases the VBW. The combination of (O-p)+(Zn-d) corrections presents a smaller VBW
than the pure LDA calculation. However, both corrections increase the value
of the energy gap. The combination of O-p and Zn-d corrections results
in energy gap in very good agreement with experiment, which again states
the importance of taking into account the cation d-state. It is worth
to point out that we obtain this good result, in spite of the fact that the
position of Zn d-state ($\sim $5 eV below the top of valence band) is higher
in energy than the experimental data ($\sim $7.8 eV). \cite{ZnO}
In the case of InN, as the d state is deeper than in the case of ZnO, the
In d-state correction is more important for the VBW value, while the N-p correction
is more important
for the energy gap value. Particularly, with (N-p)+(In-d) correction we obtain a value
for the energy gap which is in good agreement with experiment. Moreover, our
full N-p + In-d LDA-1/2 calculation is in precise agreement with the measured
value obtained from x-ray photoemission spectroscopy experiments \cite{Bechstedt},
from which the d state of In atom is found to lie 16.0 eV below the valence
band maximum.

In order to analyze the band dispersion in more detail, we also performed
calculations to obtain the conduction band effective masses. Thus, now we focus
our attention on the electronic structure around
the conduction-band minima.
We fit  a parabola to the curves of energy
versus $\mathbf{k}$ around the conduction-band minimum up to $1.0$\% along the
main symmetry directions of the Brillouin Zone. Considering the degeneracies
and making weighted averages we obtain the electron effective masses.
Table \ref{massas} summarizes the effective conduction band masses
for several semiconductors. Since a negative value for
 the LDA-InN band gap was obtained, it was not possible  to calculate an
effective mass in that case and only the LDA-1/2 value is shown. We note from
the Table that the LDA-1/2 method systematically gives larger
electron effective masses than LDA. This is due to the fact that with the
LDA underestimation of band gap energy, the $\vec{k}\cdot \vec{p}$
interaction between valence (VB) and conduction (CB) bands is stronger leading to
smaller effective masses.
Therefore, in the cases where the correction of the energy gap is more
pronounced, the difference between the LDA and LDA-1/2 electron
effective masses is larger. This is the case as, e.g., the GaAs and
ZnO, for which the LDA values agree rather poorly with experimental
data, and the LDA-1/2 gives excellent agreement with experiment. Moreover,
if we take a look at the whole table we observe that the LDA-1/2 effective
masses are generally in very good agreement with experimental data.

Therefore, the LDA-1/2 not only improves the band gaps as a ``scissor operator''
approach, but also provides reliable important band structure-derived properties,
such as the effective masses.

\section{Summary}

The very important problem concerning the calculation of excitations in solids is
addressed and a method to overcome this problem is 
developed. The method is inspired in the simple
 half-ionization method. The localization of the hole created by
 promoting an electron
from the valence band to the conduction band follows naturally from
the method. The hole is shown to be
representable by a square integrable wavefunction, instead of the usual
Bloch wave hole of band structure calculations.  

The major success of this method is its reliable description of excited states
in solids, giving band gap energies, effective masses and band structures in very
good agreement with experiment, even in the cases for which the LDA markedly
fails, e.g., ZnO and InN. The method is not more computationally demanding than
the LDA calculations.  Moreover, the method is general and can be applied to a
broad class of DFT self-consistent methods, all-electron and
pseudopotential based.

\section{Acknowledgment}
 This work was supported by the Brazilian funding
agencies FAPESP  (procs. 2006/05858-0 and 2006/61448-5)
and CNPq.

\begin{table}[htbp]
\begin{small}
    \centering
    \caption { Band energy gaps (eV) for several semiconductors
obtained with the LDA-1/2 at experimental lattice constant,
 by using the VASP code and SIESTA (S), compared
with pure LDA, GW  and experimental results of Ref. \onlinecite{Vurgaftman},
 except where noted. Direct energy gaps are noted as (d) and indirect as (i). 
The majority of the LDA-1/2 calculations were obtained {using only
the trimmed self-energy potential of p-anion}, exceptions are noted}
    \label{results}
        \begin{tabular}{lllll}
            \hline
  \hline
	 	& LDA-1/2 		& LDA & Exp. & GW \\ \hline

C (i)	&  5.25(S)$^*$   & 4.13  & 5.47$^{a}$ &5.48-5.77\cite{Schilfgaarde}  \\
C (d)    &  6.75(S)$^*$      & 5.54 & 7.3$^{a}$ & \\
Si (i)   &  1.137, 1.21(S) 		   &   0.51 & 1.17$^{a}$ & 1.32\cite{Zhu},
0.95-1.10\cite{Schilfgaarde}\\
Si (d)   &  2.9, 2.94(S)    &   2.54 & 3.05, 3.40$^{a}$  &\\
Ge (i)	&  0.70		   & 0.08	& 0.66-0.74$^{a}$&0.66-0.83\cite{Schilfgaarde} \\
AlN (d)   	&   6.06 	& 4.27  &   6.23 &5.83-6.24\cite{Schilfgaarde}\\
GaN (d) 	&    3.52$^{**}$ 				& 1.95 &   3.507 &3.15-3.47\cite{Schilfgaarde}\\
InN (d)	&   0.95 $^{**}$				&  -0.29 &  0.7-1.9 &0.20-0.33\cite{Schilfgaarde}\\
AlP (i)		&  2.79					&   1.47 & 2.52 & 2.59\cite{Zhu}\\
GaP (i)	&    2.36($\Gamma$-L)$^{**}$  & 1.49($\Gamma$-X)	&  2.35 & 2.55\cite{Zhu}\\
InP (d)    &      1.12$^{**}$			&  0.50 &        1.42 & 1.44\cite{Zhu}\\
AlAs (i)	& 2.73   			   &   1.34 & 2.24     & 2.15\cite{Zhu}\\
GaAs (d)	& 1.41   			 &   0.41 &  1.519 & 1.22\cite{Zhu},
1.40-1.70\cite{Schilfgaarde}\\
InAs (d)	& 0.75				 & -0.34 & 0.417	& 0.31\cite{Zhu}	\\
ZnO (d)  	&      3.29$^{**}$ &	 0.83 &   3.4$^{a}$ &2.51-3.07\cite{Schilfgaarde} \\
ZnS (d) 	&       3.68$^{**}$&	  2.02 &   3.91$^{a}$ &3.21-3.57\cite{Schilfgaarde} \\
\hline
  \hline
    \end{tabular}
\begin{flushleft}
$^{a}$ Ref. \onlinecite{Landolt} \\
$^*$ -1/4p-1/4s \\
$^{**}$ -1/2p-anion-1/2d-cation
\end{flushleft}
    \end{small}
\end{table}

\begin{table}[htbp]
\begin{small}
    \centering
    \caption { InN and ZnO valence band width (VBW) and band gap energy values 
     at different levels of the LDA-1/2 calculation procedure.
     The levels presented are: (i) standard LDA 
    calculation, (ii) LDA-1/2 anion-p correction only, (iii)  LDA-1/2 cation-d 
correction only, 
and    (iv) full LDA-1/2 anion-p + cation-d corrections.}
    \label{VBW}
        \begin{tabular}{lccc}
            \hline
  \hline
& correction & VBW (eV) & band gap (eV)\\ \hline

ZnO &   none&  17.72& 0.83 \\
  &O-p &   19.44  &    2.14\\
 &   Zn-d &17.01&1.48\\
 &  (O-p)+(Zn-d)&17.28&3.29\\
 InN &   none &15.25&-0.29\\
 &  N-p&14.49 &1.16\\
 & In-d&18.23  &-0.49\\
 & (N-p)+(In-d)&16.85&0.95\\
 
\hline
  \hline
    \end{tabular}
    \end{small}
\end{table}

\begin{table}[htbp]
\begin{small}
    \centering
    \caption { Effective masses (units of electron free mass, $m_e$),
for several semiconductors
obtained with the LDA-1/2 at experimental lattice constant,
 compared with pure LDA and experimental results.
 The calculations were made using the VASP code. The experimental data were 
extracted from Ref. \onlinecite{Vurgaftman},
 except where noted. { The same trimmed self-energy potential used in Table \ref{results} were also used here.}}
    \label{massas}
        \begin{tabular}{lccc}
            \hline
  \hline
&\multicolumn{3}{c}{Electron Effective mass}\\
& LDA-1/2 & LDA & Exp.\\ \hline

AlN &   0.38&0.30&-\\
GaN &   0.30&0.17&0.18-0.29\\
InN &   0.12&-&0.11-0.23\\
AlP &  0.256&0.18&-\\
GaP &   0.17&0.10&0.09-0.17\\
InP &   0.088&0.04\ \ \ \ \ &0.077-0.081\\
AlAs& 0.064&0.022&0.06-0.15\\
GaAs & 0.064&0.026&0.065-0.07\\
InAs & 0.047&0.033&0.023-0.03\\
ZnO  &  0.39&0.14&0.3-0.36\cite{ZnO_massa}\\
ZnS  &  0.26&0.16&\\
\hline
  \hline
    \end{tabular}
    \end{small}
\end{table}

\newpage

\end{document}